\documentstyle[12pt]{article}
\newcommand{\beq}{\begin{equation}}
\newcommand{\eeq}{\end{equation}}
\newcommand{\bea}{\begin{eqnarray}}
\newcommand{\eea}{\end{eqnarray}}

\newcommand{\oG}{\stackrel{o}{\Gamma}}
\begin{document}
\topmargin 0pt
\oddsidemargin 5mm
\renewcommand{\thefootnote}{\fnsymbol{footnote}}
\newpage
\setcounter{page}{0}
\begin{titlepage}
\begin{flushright} 
{\tt DIAS-STP-99-03\\ 
     hep-th/9904119
}
 \end{flushright}
\bigskip
\bigskip
 
\begin{center}
{\Large Renormalization group flow and parallel transport with non-metric compatible connections}
\bigskip 
\bigskip

{Brian P. Dolan\footnote{Department of Mathematical Physics, National University of Ireland,
	Maynooth, Republic of Ireland
	{\it and}\hskip 0.5cm Dublin Institute for Advanced Studies, 10 Burlington Rd.,
	Dublin 4, Republic of Ireland; \hfill\break e-mail: bdolan@thphys.may.ie} and
 Alex Lewis\footnote{Department of Mathematical Physics, National University of Ireland,
	Maynooth, Republic of Ireland; e-mail: alex@thphys.may.ie, supported by Enterprise Ireland grant no. SC/98/739}.}
\end{center}
\begin{center}
\footnotesize
        
	\end{center}               

\normalsize 
\bigskip 
\bigskip
\begin{center}
			{\bf Abstract}
\end{center}
A family of connections on the space of 
couplings for a renormalizable field theory
is defined. The connections are obtained 
from a Levi-Civita connection, for a metric which
is a generalisation of the Zamolodchikov metric in two 
dimensions, by adding a family
of tensors which are solutions of the renormalization group equation
for the operator 
product expansion co-efficients. The connections are 
torsion free, but not metric
compatible in general. The renormalization group 
flows of $N=2$ supersymmetric Yang-Mills
theory in four dimensions and the $O(N)$-model in 
three dimensions, in the large $N$ limit,
are analysed in terms of parallel transport under 
these connections.
\end{titlepage}
\newpage


In this letter we investigate geometrical properties of the
renormalization group flow in some exactly solved theories. 
The renormalization group
flow can be seen as a vector flow in the space of theories, with the
couplings of the theory $g^a$ being coordinates on this space . In
this approach, it has been shown in \cite{dolan1} following a
suggestion in \cite{jo} (see also \cite{ocs,sonoda} that the
renormalization group equations for multi-point correlation 
functions, written in a
coordinate covariant form, depend on a 
symmetric connection $\Gamma^a_{bc}$ through
a tensor $\tau^a_{bc}$,
\beq
\tau^a_{bc} = \nabla_b \nabla_c \beta^a - R^a_{cbd}\beta^d,
\label{tauabc}\eeq
defined by 
the RG equation for a regularized 3-point function
$G_{abc}(p,q,r)=\langle\Phi_a(p)\Phi_b(q)\Phi_c(r)\rangle$ 
\beq
\left(\Lambda\frac{\partial}{\partial\Lambda} +{\cal L}_\beta\right) 
G_{abc}(p,q,r)=\tau_{ab}^dG_{dc}(p+q,r) + \tau_{bc}^d
G_{da}(q+r,p) + \tau^d_{ac}G_{db}(r+p,q) + \cdots
\label{rgeqn}\eeq
where $G_{ab}(p,q)=\langle\Phi_a(p)\Phi_b(q)\rangle$ and the dots denote
contact terms that are only important for large momenta.
However, there is no general rule for finding a
connection. Moreover, since the RG equations only depend on the
connection through the tensor $\tau^a_{bc}$, there is in fact a family
of connections which give the same equations. The approach we take is
therefore to determine the full family of possible connections for
some exactly solvable models, and investigate the geometrical
properties of the RG flow for the most general connection. 
 For two connections $\Gamma$ and $\tilde\Gamma$ with
covariant derivatives $\nabla$ and $\tilde\nabla$ and curvatures $R$
and $\tilde R$ respectively to both be compatible with eq. (\ref{tauabc})
 we must have
\beq
\nabla_b \nabla_c \beta^a - R^a_{cbd}\beta^d =
\tilde\nabla_b \tilde\nabla_c \beta^a - \tilde{R}^a_{cbd}\beta^d.
\eeq
This equation is satisfied if the Lie derivative ${\cal L}_\beta$ of
the difference between the connections vanishes. This enables us to
determine the full family of possible connections if 
one connection
$\oG^a_{bc}$ is already known: we can 
write any connection which is compatible with eq. (\ref{tauabc})
as $\Gamma^a_{bc}=
\oG^a_{bc} + {\cal G}^a_{bc}$, where
\beq
{\cal L}_\beta  {\cal G}^a_{bc} =
{\cal G}^a_{dc}\partial_b\beta^d +
{\cal G}^a_{bd}\partial_c\beta^d -
{\cal G}^d_{bc}\partial_d\beta^a +
\beta^d\partial_d{\cal G}^a_{bc} =0.
\label{Liedv}\eeq
We still have to find a connection $\oG^a_{bc}$ to construct the
other possible connections. One solution is to use the Levi-Civita
connection of a metric on the space of couplings. An example of 
such a metric 
 is the
Zamolodchikov metric in $D=2$, which was used in the proof of the
$c$-theorem \cite{zamolodchikov}. More recently, building on ideas
laid out in \cite{ocs}, the geometrical
properties of metrics in $D>2$ have also 
been investigated for some models,
including: free field theory \cite{dolan2},
the O(N) model \cite{dolan3} and Seiberg-Witten theory for $SU(2)$
\cite{dolan4}. In all these models, it was found that
some (but not all) of the RG flow lines are geodesics of the
metric. In particular, the lines of crossover between fixed points are
geodesics, and this may be related to irreversibility of the
renormalization group flow.
 Since we now have  a family of connections which
are equivalent, at least as far 
as the renormalization group equations are concerned,
 it is natural to ask whether the geodesic
flow of the lines of crossover generalises to auto-parallel flow for
other connections 
 (a line which is auto-parallel for the Levi-Civita
connection is a geodesic)\footnote{We distinguish between
auto-parallels, which are curves whose tangent vectors remain tangent
vectors under parallel transport along the curve, and geodesics, which
are curves of shortest length. In general, these coincide for the
 Levi-Civita
connection only.}.

The auto-parallel
equation for a vector field is $\nabla_{\vec\beta}
\vec\beta = \eta\vec\beta$, where $\eta$ is a function which depends
on the parameterization along the curve.
   With a connection $\Gamma=\oG +
{\cal G}$, this becomes
\beq
\beta^b\frac{\partial\beta^a}{\partial x^b} +
\oG^a_{bc}\beta^b\beta^c + {\cal G}^a_{bc}\beta^b\beta^c =
\eta\beta^a. 
\label{fulleq}\eeq
Our main aim is to see which of the possible connections, if any,
 will satisfy
 this equation for a given renormalization group trajectory. In
 particular, some of the trajectories in the models we will examine in
 this letter are geodesics of the metric, so  
  eq. (\ref{fulleq}) is satisfied 
for ${\cal G}=0$. In that case, eq. (\ref{fulleq}) simplifies to the
 condition that
\beq
{\cal G}^a_{bc}\beta^b\beta^c =
\eta'\beta^a, 
\label{auto}\eeq
where $\eta'$ is another function.

A natural candidate for a metric on the space of couplings is the
two-point correlation functions of the model \cite{ocs}. If 
 the action $S$ is
linear in the couplings,
\beq
S = S_0  + \int d^Dx g^a\Phi_a(x)
\eeq
then a metric can be defined by
\beq
G_{ab} = \int d^Dx \langle \tilde\Phi(x)\tilde\Phi(0) \rangle
\label{metric}\eeq
where $\tilde\Phi(x)=\Phi(x)-\langle\Phi(x)\rangle$. Although the
individual components of this metric may diverge, the geometry can
still be non-singular.

As our first example, we 
consider the $O(N)$ model for large $N$ in 3 dimensions.
This is a model of a scalar field $\vec\varphi$ in 
the vector representation
of $O(N)$ with the action
\beq
S=\int d^3x \left\{ \frac{1}{2} (\nabla\vec\varphi)^2
+\vec{j}\cdot\vec\varphi +\frac{r}{2}\vec\varphi^2
+\frac{u}{4!}(\vec\varphi^2)^2 
\right\}
\eeq
following \cite{dolan3} we analyse the geometry in terms of three
 bare parameters, $\phi ,X,\lambda$, defined by
\beq
\phi = 4\sqrt\frac{\pi }{N\Lambda}\langle\varphi\rangle,
~~~~~X = \frac{1}{2\Lambda}\int d^3x \langle\varphi^2\rangle,
~~~~~\lambda = \frac{Nu}{48\pi\Lambda}.
\eeq
Although these are bare parameters, they are finite as we have a
finite cut-off $\Lambda$, so we can use them 
as our coordinates on the space of
couplings (since we are in any case only interested in properties of
the RG flow which are independent of the coordinate system). The beta
functions, which represent a vector flow on this space are
\cite{dolan3}
\beq
\beta^\phi=-\frac{1}{2}\phi~~~~~\beta^X=-X~~~~~\beta^\lambda =-\lambda.
\label{betas}\eeq
In \cite{dolan3} the metric (\ref{metric})
was computed, and
it was found that only one of the renormalization group 
trajectories described by these
beta functions is actually a geodesic of the metric - the line
$X=\phi=0$, which is the line of crossover from the Wilson-Fisher
fixed point at $\lambda=\infty$ to the Gaussian fixed point at
$\lambda=0$.
 We now want 
to see if any of the renormalization group trajectories are
auto-parallel for a connection $\Gamma=\oG +{\cal G}$, where $\oG$ is
the Levi-Civita connection from \cite{dolan3} and ${\cal G}$ is a
solution of eq. (\ref{Liedv}). Using bare rather than renormalized
parameters as coordinates makes it easy to solve eq. (\ref{Liedv}) --
the general solution contains a number of arbitrary functions
 which
depend only on the ratios 
$X/\lambda$ and $\phi^2/\lambda$, but these ratios are constant along
any of the RG trajectories described by eq. (\ref{betas}), so we can
treat them as just being arbitrary constants, which we write as
$f^i_{jk}$.
The solution can then be written, provided $X$, $\phi$ and
$\lambda$ are non-zero, as 
\beq
{\cal G}^i_{jk} =
\frac{\beta^i}{\beta^j\beta^k}f^{i}_{jk}
\label{ONG}\eeq
(no summation on $i$, $j$, $k$).
With this solution, it is clear that for eq. (\ref{fulleq}) to be
  satisfied for a trajectory, 
the following differences must be constant along the curve: 
\beq
\frac{2}{\phi}\oG^\phi_{ij}\beta^i\beta^j-
\frac{1}{\lambda}\oG^\lambda_{ij}\beta^i\beta^j=\mbox{constant},
~~~~~~~~\frac{2}{\phi}\oG^\phi_{ij}\beta^i\beta^j-
\frac{1}{X}\oG^X_{ij}\beta^i\beta^j=\mbox{constant}.
\label{cond}\eeq
It can be seen from the
 expressions for the components of the Levi-Civita connection given in
 Appendix 2 of \cite{dolan2} that this is not true for any of the RG
 flow lines except the line of crossover. 
Thus none of the flow lines with $\phi$, $X$, and
 $\lambda$ all non-zero can be auto-parallel for
 any connection.
 When one
  of $X$, $\phi$ or $\lambda$ is $0$, the solution (\ref{ONG}) has to
  be changed by absorbing factors of $X/\lambda$ or $\phi^2/\lambda$
  into $f^i_{jk}$ to make it finite, for example if $\lambda\neq 0$ we
 can write the solution as
\beq 
{\cal G}^i_{jk} = \lambda^nf^i_{jk}~~~~~~~~n=n_i-n_j-n_k
\label{secondG}\eeq
 with $n_\phi=\frac12$ and $n_X=n_\lambda=1$.
However, if $X=0$, $\phi=0$ or $\lambda=0$,
 eq. (\ref{fulleq}) can only
be satisfied if $\oG^X_{ij}\beta^i\beta^j=0$, 
$\oG^\phi_{ij}\beta^i\beta^j=0$ or
$\oG^\lambda_{ij}\beta^i\beta^j=0$ respectively, and the only line
which satisfies these conditions is $X=\phi=0$. Thus none of the other
RG
flow lines can ever be auto-parallel.
Finally, we want to see if the geodesic $X=\phi=0$ is
 auto-parallel for other connections (apart from the Levi-Civita
connection).
This means we have to see if eq. (\ref{auto}) is satisfied by
the solution (\ref{secondG}).
 For example, if only
$f^\phi_{\phi\phi}$, $f^\lambda_{\lambda\lambda}$ and $f^X_{XX}$ are
non-zero, eq. (\ref{auto}) becomes
\bea
-\eta' \phi &=& \lambda^{-1/2}f^\phi_{\phi\phi}\phi^2/2 \nonumber \\
-\eta' \lambda &=& \lambda^{-1}f^\lambda_{\lambda\lambda}\lambda^2 \\
-\eta' X &=& \lambda^{-1}f^X_{XX}X^2 \nonumber
\eea
which clearly is satisfied for $X=\phi=0$.
However this is not true for the most general connection.
For example, if ${\cal G}^X_{\lambda\lambda}=\lambda^{-1} 
f^X_{\lambda\lambda}\ne 0$
or ${\cal G}^\phi_{\lambda\lambda}=\lambda^{-3/2} 
f^\phi_{\lambda\lambda}\ne 0$
then the line of crossover is not an auto-parallel (since the 
corresponding Levi-Civita
connection components vanish in the large $N$ limit, \cite{dolan3}).
The line of crossover is therefore not auto-parallel for the entire family
of connections which can be used in the RG equations, but only for the 
class
with ${\cal G}^X_{\lambda\lambda}={\cal G}^\phi_{\lambda\lambda}=0$.

Thus 
we find that
introducing the family of non-metric compatible connections 
changes the conclusions, compared to the the Levi-Civita case, 
as to which renormalization group
flow lines are auto-parallel. There is a large class of connections
for which the line of crossover between the Gaussian and the Wilson-Fisher
fixed points remains auto-parallel, but this is not true of the most general
connection.

Our second example of an RG flow shows that, in general, the family of 
connections do change the auto-parallel nature of the geodesic flows
(in the special cases where RG flow is geodesic).
This is 4-dimensional $N=2$ Yang-Mills
theory \cite{sw}. The geometrical properties of this model were
investigated in \cite{dolan4} where two possible metrics were
considered, 
and it was found most of the renormalization group flow lines are not
geodesics, but some special lines are geodesics of both metrics. We
now want to see which, if any, of the RG flow lines are auto-parallel
for other connections. 
The complex coupling $\tau = \frac{\theta_{eff}}{2\pi} +\frac{4\pi
i}{g^2_{eff}}$ is given as a function of $\tilde{u}=u/\Lambda^2$,
where $u=Tr\langle\varphi^2\rangle$ parameterizes the symmetry breaking,
by
\beq
\tau = \frac{iK'}{K}+2n
\label{tau}\eeq
where $K(k^2)$ is a standard elliptic integral \cite{ww}
 with $k^2 =
\frac{2}{\tilde{u}+1}$ and $K'=K(1-k^2)$ and $n$ is an 
integer. Various beta functions for
 this model have been investigated in \cite{bm,ritz,ckmm}. In
\cite{ckmm} Wilsonian and Novikov, Shifman, Vainshtein,  
Zakharov, \cite{NSVZ}, beta functions were considered. Here we
concentrate on the Callan-Symanzik beta functions of \cite{bm,ritz}.
The Callan-Symanzik
beta functions,
$\vec\beta = \beta\frac{\partial}{\partial\tau}
+\bar\beta\frac{\partial}{\partial\bar\tau}$ are defined by  
\beq
\beta(\tau) = \lambda
\left.\frac{\partial\tau}{\partial\Lambda}\right|_u =
-2\tilde{u}\frac{d\tau}{d\tilde{u}}\; .
\label{betatau}\eeq
This beta function represents a vector flow on a manifold,
parameterized by $\tau$ (or $u$), which has the topology of a sphere
with three punctures. This manifold has three singular points:
$u=\infty$ (the weak coupling limit) and $u=\pm \Lambda^2$ (where there
are extra massless degrees of freedom).
The Seiberg-Witten metric on this manifold in the $u$-coordinates is
\beq
ds^2=\pi^2 \mbox{Im}(\tau) \left|
\frac{\vartheta_3^4\vartheta_4^4}{\vartheta_2^2}\right|^2 
d\tau d\bar\tau =
\frac{1}{\pi^2} \frac{(K'\bar{K} +
\bar{K}'K)}{\sqrt{1+u}\sqrt{1+\bar{u}}}dud\bar{u},
\label{swmetric}\eeq
where $\vartheta_i$, $i=2,3,4$ are Jacobi $\vartheta$-functions, 
\cite{ww}.

Another metric which can be introduced for this geometry is the
Poincare metric
\beq
ds^2=\frac{1}{(\mbox{Im}(\tau))^2}d\tau d\bar\tau.
\eeq
For both these metrics, the lines 
of real $u$ and imaginary $u$ are
geodesics, but the other RG flow lines are not
\cite{dolan4}. 

In fact we do not need to know the explicit form of $\beta(\tau)$ to
see if the geodesics will be auto-parallel for any connection: if
we use $u$ and $\bar u$ as coordinates, so that $\vec\beta = \beta
(u)\frac{\partial}{\partial u} +\bar\beta (u) \frac{\partial}{\partial
\bar{u}}$, we  only need to know that
$\frac{\partial\beta}{\partial \bar
u}=\frac{\partial\bar\beta}{\partial u}= 0$.
 The solution
of eq. (\ref{Liedv}) is then
\bea
{\cal G}^u_{uu} = \frac{1}{\beta}g^u_{uu}(u/\bar{u})
&& 
{\cal G}^{\bar u}_{\bar u\bar u} =
\frac{1}{\bar\beta}g^{\bar u}_{\bar{u}\bar{u}}(u/\bar{u})
\nonumber\\ 
{\cal G}^u_{\bar{u}u} =
\frac{1}{\bar\beta}g^u_{\bar{u}u}(u/\bar{u}) 
&&
{\cal G}^{\bar u}_{u\bar{u}} =
\frac{1}{\beta}g^{\bar u}_{u\bar{u}}(u/\bar{u}) \nonumber \\
{\cal G}^u_{\bar u\bar u} =
\frac{\beta}{\bar\beta^2}g^u_{\bar{u}\bar{u}}(u/\bar{u})
&&
{\cal G}^{\bar u}{uu} = 
\frac{\bar\beta}{\beta^2}g^{\bar u}_{uu}(u/\bar{u}).
\eea
The solution contains arbitrary functions of $u/\bar{u}$, 
but these functions are
 constant
along the radial lines in the $u$-plane, and we can see from
 eq. (\ref{betatau}) that these are just the RG flow lines (in the
 $u$-coordinates, $\beta^u=-2u$, $\beta^{\bar u}=-2\bar u$). If we
assume that ${\cal G} = {\cal G}^i_{jk}dx^j\otimes
 dx^k\otimes\frac{\partial}{\partial x^i}$ is a real tensor and
  substitute this solution into eq. (\ref{fulleq}) or (\ref{auto}), we
 can see that those equations depend only on one function $g(u/\bar
 u)$:
\beq
g(u/\bar{u}) = g^u_{uu}+g^u_{\bar u u}+g^u_{u\bar u}+g^u_{\bar u\bar
u} ~~~~~ \bar{g}(u/\bar{u}) = g^{\bar u}_{\bar u\bar
u}+g^{\bar u}_{u\bar u}+g^{\bar u}_{\bar u u}+g^{\bar u}_{uu}.
\eeq 
Eq. (\ref{fulleq}) then reduces to the condition that
\beq
\oG^u_{uu} u +g(u/\bar u) = \oG^{\bar u}_{\bar u\bar u} \bar u +\bar
g(u/\bar u) 
\eeq
where for the Seiberg-Witten metric $\oG^{\bar u}_{\bar u\bar u} =
\overline{{\oG}^u_{uu}}$ and 
\beq
\oG^u_{uu} = \frac{\bar{K}'\partial_u K
+\bar{K}\partial_uK'}{K'\bar{K} +\bar{K}'K} -\frac{1}{2(u+1)}.
\eeq
Since $g$ and $\bar g$ are constant along an RG trajectory,
 a trajectory can only be
auto-parallel for some choice of $g$ and $\bar g$ if the imaginary part
of $\oG^u_{uu}u$ is constant along that trajectory. It can be
shown numerically that this
is only the case for the geodesics, the lines of real and imaginary
$u$, where $\oG^u_{uu}u$ is real. All the other RG flow lines are
therefore not auto-parallel for any connection, while the real $u$ line
is auto-parallel provided $g(1)=\bar{g}(1)$ and the imaginary $u$ line
is auto-parallel provided $g(-1)=\bar{g}(-1)$. Thus 
the geodesic flow lines
are not auto-parallel for the most general connection (as was the case
in the $O(N)$ model), but they are auto-parallel, for example, if
 $\bar{g}(u/\bar u)= g(\bar u/u)$.

Since we did not use the explicit expressions for $\tau$ and $\beta$
above, this result can be applied to any theory in which the beta
function is an analytic function of one complex variable
only. In the case of $N=2$ $SU(2)$
Yang-Mills theory with massless quarks, as in the pure
Yang-Mills theory, the complex coupling $\tau$ depends only on
$u/\Lambda^2$. However, the singularities of the theories with
massless quarks are not all on the real or imaginary $u$ axes - for
example, when the number of flavours $N_f=1$, there are singularities
at $u=-u_0$ and $u=u_0e^{\pm i\pi/3}$. Of course, we cannot tell if 
these lines are actually
auto-parallel unless we know whether they are geodesics of the
Seiberg-Witten
metric, but we can say that if they are geodesics of the
Seiberg-Witten metric in these models, they be auto-parallel
for  connections with $g(e^{2\pi i/3})=\bar {g}(e^{2\pi i/3})$,
 but not for all connections for which 
$\bar{g}(u/\bar u)= g(\bar u/u)$, as in the case of the pure Yang-Mills
theory.

In conclusion it has been shown that the renormalization group
equation for the operator expansion co-efficients gives rise
to a family of non-metric compatible connections which are related
to the Levi-Civita connection by 
$\Gamma^a_{bc}=
\oG^a_{bc} + {\cal G}^a_{bc}$, where
${\cal L}_\beta  {\cal G}^a_{bc} =0$.
In general RG flows which are geodesic for the Levi-Civita connection
are not auto-parallel for all members of the family,
though in the models examined they are for a large sub-class of 
connections
(it can be shown that 
the same is also true for the 
free field models considered in \cite{dolan2}).
In none of the examples examined here is it the case that a RG flow
line which is not geodesic under the Levi-Civita connection
is auto-parallel for some member of the family.

\end{document}